\documentclass[runningheads]{llncs}
\usepackage[utf8]{inputenc}
\usepackage[T1]{fontenc}

\usepackage[english]{babel}

\usepackage{isabelle,isabellesym}
\isabellestyle{tt}

\usepackage{todonotes}

\usepackage{graphicx}
\usepackage{amsmath,amsfonts}

\usepackage[most,breakable]{tcolorbox}

\newtcolorbox{isaframe}[1][]
   { blanker, 
    left=5pt, right=1pt, top=1pt, bottom=1pt,
     borderline west={1pt}{0pt}{orange},
     before upper=\setlength{\parindent}{0pt},
     fontupper=\ttfamily,
     parbox=true, #1}

\usepackage[draft=false,breaklinks,colorlinks]{hyperref}

\usepackage[nameinlink]{cleveref}
\crefname{theorem}{theorem}{theorems}
\crefname{corollary}{corollary}{corollaries}
\crefname{example}{example}{examples}
\crefname{lemma}{lemma}{lemmas}
\crefname{proposition}{proposition}{propositions}
\crefname{definition}{definition}{definitions}
\crefname{observation}{observation}{observations}

\newcommand{\isaterm}[1]{\texttt{#1}} 

\newcommand\rev[1]{%
\ensuremath{\mathop\mathrm{rev}%
\ifx&#1&%
\else
  {\left( #1 \right)}
\fi
}}

\newcommand{\atype}{\isaterm{{\isacharprime}a}}
\newcommand{\alist}{\isaterm{{\isacharprime}a\ list}}
\newcommand{\blist}{\isaterm{{\isacharprime}b\ list}}
\newcommand{\astream}{\isaterm{{\isacharprime}a\ stream}}

\newcommand{\blistlist}{\isaterm{{\isacharprime}b\ list list}}
\newcommand{\alistset}{\isaterm{{\isacharprime}a\ list set}}

\newcommand{\Nil}{\texttt{Nil}}

\newcommand{\A}{\mathcal{A}}

\newcommand{\uu}{\mathbf u}
\newcommand{\vv}{\mathbf v}

\begin{document}
\title{Infinite Words and Morphic Languages Formalized in Isabelle/HOL}
%
%
\author{
Štěpán Starosta\inst{2}\orcidID{0000-0001-5962-4297}
}
\authorrunning{Š. Starosta}
%
\institute{
Czech Technical University in Prague, Czech Republic\\
}


%
\maketitle              
\begin{abstract}
We present a formalization of basics related to infinite words in the generic proof assistant Isabelle/HOL.
Furthermore, we present a formalization of purely morphic and morphic languages.
Finally, we present a formalized definition of Sturmian words as lower mechanical words and prove some very elementary facts.
The formalization is based on an ongoing larger project of formalization of combinatorics on words.
\end{abstract}

\section{Introduction}

Infinite words and their various properties are a popular object of study in combinatorics on words.
While the beginning of such studies is usually attributed to M. Prouhet's work \cite{Prouhet} of 1851 and A. Thue's work \cite{Thue06} some 50 years later, the beginning of a systematic study and raise of the field is marked by the creation of the first of Lothaire's book \cite{Lo83}.
For a detailed description of the history and on the studied concepts one may refer to \cite{BePe}.

This article presents a formalization of elementary concepts related to infinite words from the point of view of combinatorics on words.
The goals of mathematics formalization is to write down statements in a formal way and provide them with machine-verified proofs using a given set of inference rules.
With the raise of computing power, this field has grown considerably since the first (large) computer-assisted proof of the four-color theorem in 1970s.
The reader may refer to the survey \cite{KaRa} for a more detailed overview of the state of mathematics formalization.
From a different angle, the evolution of this field and its available tools may be also tracked on the formalization state of 100 prominent theorems in various available provers \cite{Freek100}.

The presented formalization is done in the generic proof assistant Isabelle/HOL \cite{Isabelle,IsabelleHOLBook}.
It is an open-source project with a vast collection of formalized results available in the main distribution of Isabelle and in the Archive of Formal Proofs \cite{afp}.
The interface language of Isabelle is called Isar, see \cite{IsarTutorial}.
One of Isar's goals is to provide human-readable proofs.
In what follows, we give many examples of Isar code copied as it is in the presented formalization, leaving to the reader to consider the achieved level of readability.

The formalization is within an existing project of formalization of combinatorics on words in Isabelle/HOL \cite{CoW_gitlab} called ``Combinatorics on Words Formalized''.
As of today, it covers many elementary and some advanced concepts from the field (related to finite words).
Besides the most up-to-date repository \cite{CoW_gitlab}, the reader may refer to \cite{itp2021,binary_codes_IJCAR,DLT_Lyndon} for more details on this project.
In this context, the present article describes the extension of the project with the concepts of infinite words and, in what follows, gives highlights of important or interesting parts of this extension, omitting numerous other formalized claims and proofs.

In the next section we give a few more details on relevant concepts from the Combinatorics on Words Formalized project.
The formalization is split into 3 parts, described \Cref{sec:languages,sec:infinite_words,sec:sturmian_words}.
The last section describes the structure of the formalization.

\section{Preliminaries - combinatorics on finite words}

The basic component of the Combinatorics on Words Formalized project, called ``Combinatorics on Words Basics'', is also available in the Archive of Formal Proofs \cite{Combinatorics_Words-AFP}, where it is shortly called ``Combinatorics\_Words''.
We shall refer to it by this name as it is the basic component which allows the formalization of the concepts presented in this article.
For the same reason, in this section, we give a very brief overview of related notions available in Combinatorics\_Words.

Finite words are finite sequences of elements from a given set $\A$, which is usually called an \emph{alphabet}.
The elements of an alphabet are \textit{letters}.
Finite words are naturally represented by the datatype of lists, denoted \alist{} in Isabelle, where \atype{} is a type variable.
Hence, \atype{} represents the type for the letter, and all the elements of this type represent the alphabet.
The empty list/word is \isaterm{{\isacharbrackleft}{\kern0pt}{\isacharbrackright}} or \Nil{}, also denoted by $\varepsilon$ in Combinatorics\_Words.

The operation of concatenation is also provided by the ``List'' theory of Isabelle main distribution.
A theory is the elementary unit of formalization in Isabelle and it coincides with a file having extension ``thy''.
Combinatorics\_Words package defines the more usual notation of concatenation in combinatorics on words, namely, it
is denoted in the formalization by the usual algebraic notation $\cdot$.
The length of a finite word is also defined in the List theory, and a usual notation $|w|$\footnote{The symbol ``|'' is in boldface in the formalization to avoid conflict with existing notation.} is adopted in Combinatorics\_Words.
Similarly, the concepts of prefix and factor are already in the theory ``HOL-Library.Sublist'' which is reused by Combinatorics\_Words.
A frequent notation for these phenomena is introduced: $p \leq_p w$ stands for $p$ being a prefix of $w$, and $f \leq_p w$ denotes that $f$ is a factor of $w$.
The mentioned notation, along with many elementary or slightly advances related claims are in ``CoWBasic'' theory of Combinatorics\_Words.

The concept of a morphism and endomorphism is introduced in ``Morphisms'' theory of Combinatorics\_Words along with many useful claims.
A morphism $\varphi$ is a mapping from a set of finite words to another set of finite words such that for all $w,v$ we have $\varphi(w \cdot v) = \varphi(w) \cdot \varphi(v)$ (it is a homomorphism from one word monoid to another).

\section{Languages} \label{sec:languages}

The study of infinite words often comprises the study of their set of factors.
Hence, naturally, before introducing infinite words, we introduce several notions relevant to languages, i.e., sets of finite words, in general.

A language in the formalization needs not be defined as it stems from the fact that the alphabet is represented by a type.
In other words, a language is a variable of the type \alistset{}, i.e., set of variables of type \alist{}, that is, a set of finite words.

The first definition is a language closed under taking factors.
Such language is called \emph{factorial}: a language $L$ is factorial if
\[
\forall w \in L, \forall f, f \leq_f w \Rightarrow f \in L.
\]
The formalization of this definition is straightforward:

\begin{isaframe}
\isacommand{definition}\isamarkupfalse%
\ factorial\ {\isacharcolon}{\kern0pt}{\isacharcolon}{\kern0pt}\ {\isachardoublequoteopen}{\isacharprime}{\kern0pt}a\ list\ set\ {\isasymRightarrow}\ bool{\isachardoublequoteclose}\isanewline
\ \ \isakeyword{where}\ {\isachardoublequoteopen}factorial\ L\ {\isasymequiv}\ {\isacharparenleft}{\kern0pt}{\isasymforall}w\ f{\isachardot}{\kern0pt}\ w\ {\isasymin}\ L\ {\isasymand}\ f\ {\isasymle}f\ w\ {\isasymlongrightarrow}\ f\ {\isasymin}\ L{\isacharparenright}{\kern0pt}{\isachardoublequoteclose}
\end{isaframe}

The type {\isachardoublequoteopen}{\isacharprime}{\kern0pt}a\ list\ set\ {\isasymRightarrow}\ bool{\isachardoublequoteclose} represents a mapping from sets of \alist, i.e., sets of finite words, to bool, thus creating a predicate on a language.
The symbol \isasymlongrightarrow{} stands for HOL implication, i.e., a binary operation acting on the bool type.

Taking the \emph{factor closure} of a language $L$ produces the smallest superset of $L$ which is factorial.
In the formalization, the command \isacommand{inductive{\isacharunderscore}{\kern0pt}set} is used to define the closure inductively:

\begin{isaframe}
\isacommand{inductive{\isacharunderscore}{\kern0pt}set}\isamarkupfalse%
\ factor{\isacharunderscore}{\kern0pt}closure{\isacharcolon}{\kern0pt}{\isacharcolon}{\kern0pt}\ {\isachardoublequoteopen}{\isacharprime}{\kern0pt}a\ list\ set\ {\isasymRightarrow}\ {\isacharprime}{\kern0pt}a\ list\ set{\isachardoublequoteclose}\ \isakeyword{for}\ L\isanewline
\ \ \isakeyword{where}\ {\isachardoublequoteopen}w\ {\isasymin}\ L\ {\isasymLongrightarrow}\ v\ {\isasymle}f\ w\ {\isasymLongrightarrow}\ v\ {\isasymin}\ factor{\isacharunderscore}{\kern0pt}closure\ L{\isachardoublequoteclose}
\end{isaframe}

The long double arrow {\isasymLongrightarrow} is Isabelle's metaimplication using to reason about claims.
The last line should be read as if $w \in L$ and $v \leq_f w$, then $v$ belongs to \isaterm{factor{\isacharunderscore}{\kern0pt}closure\ L}.
The free variables \isaterm{w} and \isaterm{v} are universally quantified.
The advantage of using this, very simple, inductive definition is that Isabelle produces many useful facts that can be used later in the formalization.
For instance, the following induction rule, called \isaterm{factor\_closure.induct} is produced:
\begin{isaframe}
\ \ \ \ {\isachardoublequoteopen}x\ {\isasymin}\ factor{\isacharunderscore}{\kern0pt}closure\ L\ {\isasymLongrightarrow}\isanewline
\ \ \ \ {\isacharparenleft}{\kern0pt}{\isasymAnd}w\ v{\isachardot}{\kern0pt}\ w\ {\isasymin}\ L\ {\isasymLongrightarrow}\ v\ {\isasymle}f\ w\ {\isasymLongrightarrow}\ P\ v{\isacharparenright}{\kern0pt}\ {\isasymLongrightarrow}\ P\ x{\isachardoublequoteclose}%
\end{isaframe}

This automatically produced rule may be read as if $x$ is in the factor closure of a language $L$ and for all $w,v$ we have that
\[
w \in L, v \leq_f w \text{ implies that } P(v) \text{ holds,}
\]
then $P(x)$ holds.



There are more advantages of using the command \isacommand{inductive{\isacharunderscore}{\kern0pt}set} when used to define the following concept of a morphic language,
which stems from the idea of a D0L-system.
A D0L-system is a specific instance of an L-system introduced in \cite{LINDENMAYER1968280} to model growth of organisms and used nowadays to generate vegetations in virtual worlds.
A classical overview of related concepts is in \cite{book-of-L}.
A D0L-system is a triple $(\A,\varphi,w)$ where $\A$ is an alphabet, $\varphi$ is an endomorphism over $\A^*$, the set of all finite words over $\A$, and $w$ is a word over $\A$ usually called \emph{axiom}.
The set $\left\{ \varphi^k (w) \mid k=0,1,\dots \right \}$ is the language of the system.
We are interested in the factorial closure of this language, that is, we are interested in any factor that can be produced by repeated application of $\varphi$ on the axiom $w$.
We call this language a \emph{purely morphic language} and it is defined as follows:

\begin{isaframe}
\isacommand{inductive{\isacharunderscore}{\kern0pt}set}\isamarkupfalse%
\ purely{\isacharunderscore}{\kern0pt}morphic{\isacharunderscore}{\kern0pt}language\ {\isacharcolon}{\kern0pt}{\isacharcolon}{\kern0pt}\ {\isachardoublequoteopen}{\isacharparenleft}{\kern0pt}{\isacharprime}{\kern0pt}a\ list\ {\isasymRightarrow}\ {\isacharprime}{\kern0pt}a\ list{\isacharparenright}{\kern0pt}\ {\isasymRightarrow}\ {\isacharprime}{\kern0pt}a\ list\ {\isasymRightarrow}\ {\isacharprime}{\kern0pt}a\ list\ set{\isachardoublequoteclose}\isanewline
\ \ \isakeyword{for}\ f\ axiom\ \isakeyword{where}\isanewline
\ \ \ \ {\isachardoublequoteopen}axiom\ {\isasymin}\ purely{\isacharunderscore}{\kern0pt}morphic{\isacharunderscore}{\kern0pt}language\ f\ axiom{\isachardoublequoteclose}\isanewline
\ \ {\isacharbar}{\kern0pt}\ {\isachardoublequoteopen}w\ {\isasymin}\ purely{\isacharunderscore}{\kern0pt}morphic{\isacharunderscore}{\kern0pt}language\ f\ axiom\ {\isasymLongrightarrow}\ f\ w\ {\isasymin}\ purely{\isacharunderscore}{\kern0pt}morphic{\isacharunderscore}{\kern0pt}language\ f\ axiom{\isachardoublequoteclose}\isanewline
\ \ {\isacharbar}{\kern0pt}\ {\isachardoublequoteopen}w\ {\isasymin}\ purely{\isacharunderscore}{\kern0pt}morphic{\isacharunderscore}{\kern0pt}language\ f\ axiom\ {\isasymLongrightarrow}\ u\ {\isasymle}f\ w\ {\isasymLongrightarrow}\ u\ {\isasymin}\ purely{\isacharunderscore}{\kern0pt}morphic{\isacharunderscore}{\kern0pt}language\ f\ axiom{\isachardoublequoteclose}
\end{isaframe}

There are two parameters in the above definition of \isaterm{purely{\isacharunderscore}{\kern0pt}morphic{\isacharunderscore}{\kern0pt}language}.
The first one, denoted \isaterm{f}, is of type \isaterm{{\isacharprime}{\kern0pt}a\ list\ {\isasymRightarrow}\ {\isacharprime}{\kern0pt}a\ list}, which is of the same type as an endomorphism, but the requirement for it to actually be an endomorphism is not present.
The reason is that the definition itself does not need this assumption, and it will be added when proving reasonable claims about this inductive set later.
The second parameter is the axiom, denoted \isaterm{axiom}.

There are 3 inductive rules:
\begin{enumerate}
\item the axiom is in the language;
\item the image by \isaterm{f} of any element of the language is in the language;
\item a factor of any element of the language is in the language.
\end{enumerate}
Considering the underlying datatype as the alphabet, it coincides with the factor closure of a language of a D0L-system for an endomorphism \isaterm{f}.
On the other hand, for a specific choice of axiom, it coincides with the set of all factors of a purely morphic infinite word, hence the name.
An infinite word $\uu$ \emph{purely morphic} if there exists an endomorphism $\varphi$ with $\varphi(\uu) = \uu$ and $\lim_{k \to +\infty}|\varphi^k(a)| = +\infty$ where $a$ is the first letter of $\uu$.
In other words $\varphi^k(a)$ constitute longer and longer prefixes of the word $\uu$.

The fixing of the parameter \isaterm{f} to be an endomorphism is done by the command \isacommand{context}, referring to the locale \isaterm{endomorphism} from the required theory ``Morphisms'' of Combinatorics\_Words.
Locales are Isabelle's tool to fix parameters to avoid their repetition in each claim.
In this context, the endomorphism variable is fixed to \isaterm{f}, and \isaterm{{\isasymL}\ w} is a shorthand for \isaterm{purely{\isacharunderscore}{\kern0pt}morphic{\isacharunderscore}{\kern0pt}language\ f\ w}.

\begin{isaframe}
\isacommand{context}\isamarkupfalse%
\ endomorphism\isanewline
\isakeyword{begin}\isanewline
\isanewline
\isacommand{definition}\isamarkupfalse%
\ bounded\ {\isacharcolon}{\kern0pt}{\isacharcolon}{\kern0pt}\ {\isachardoublequoteopen}{\isacharprime}{\kern0pt}a\ list\ {\isasymRightarrow}\ bool{\isachardoublequoteclose}\isanewline
\ \ \isakeyword{where}\ {\isachardoublequoteopen}bounded\ w\ {\isasymequiv}\ {\isacharparenleft}{\kern0pt}finite\ {\isacharparenleft}{\kern0pt}{\isasymL}\ w{\isacharparenright}{\kern0pt}{\isacharparenright}{\kern0pt}{\isachardoublequoteclose}
\end{isaframe}

The definition reads that a word $w$ is \emph{bounded} if the purely morphic languages induced by it is finite.
The bounded words are interesting in the context when there are infinitely many bounded elements in a purely morphic language.
Such systems are in \cite{EHRENFEUCHT198313} called \emph{pushy}, which is the term we also adopt (\isaterm{f} is still fixed to be an endomorphism):



\begin{isaframe}
\isacommand{definition}\isamarkupfalse%
\ pushy\ {\isacharcolon}{\kern0pt}{\isacharcolon}{\kern0pt}\ {\isachardoublequoteopen}{\isacharprime}{\kern0pt}a\ list\ {\isasymRightarrow}\ bool{\isachardoublequoteclose}\isanewline
\ \ \isakeyword{where}\ {\isachardoublequoteopen}pushy\ axiom\ {\isasymequiv}\ {\isasymnot}\ finite\ {\isacharparenleft}{\kern0pt}{\isacharparenleft}{\kern0pt}{\isasymL}\ axiom{\isacharparenright}{\kern0pt}\ {\isasyminter}\ {\isacharparenleft}{\kern0pt}Collect\ bounded{\isacharparenright}{\kern0pt}{\isacharparenright}{\kern0pt}{\isachardoublequoteclose}
\end{isaframe}

As a highlight of a formalized elementary claim, we choose the following one stating that if an axiom is pushy with respect to the fixed endomorphism \isaterm{f} if and only if it is pushy with respect to any non-trivial power of \isaterm{f}:

\begin{isaframe}
\isacommand{lemma} pushy{\isacharunderscore}{\kern0pt}strong{\isacharunderscore}{\kern0pt}shift{\isacharcolon}{\kern0pt}\ \isanewline
\ \ \isakeyword{assumes}\ {\isachardoublequoteopen}finite\ {\isacharparenleft}{\kern0pt}UNIV{\isacharcolon}{\kern0pt}{\isacharcolon}{\kern0pt}{\isacharprime}{\kern0pt}a\ set{\isacharparenright}{\kern0pt}{\isachardoublequoteclose}\ \isanewline
\ \ \isakeyword{shows}\ {\isachardoublequoteopen}pushy\ axiom\ {\isacharequal}{\kern0pt}\ endomorphism{\isachardot}{\kern0pt}pushy\ {\isacharparenleft}{\kern0pt}f{\isacharcircum}{\kern0pt}{\isacharcircum}{\kern0pt}(p + 1){\isacharparenright}{\kern0pt}\ axiom{\isachardoublequoteclose}
\end{isaframe}

The assumption of the last lemma states that the universe of the datatype \atype{} representing the alphabet is finite.

While a proof is quite trivial for a non-erasing endomorphism \isaterm{f}, it needs to be more elaborate when there is no such assumption.
A good formalization forces the author not only to write down each claim correctly in the given formal language, 
but to choose (again and again) one of the many equivalent formulations which is the best for further reuse, as typically, most of the claims are be reused later in further formalization.
This choice comes with the revision of the assumptions which often ends in dropping or/and generalizing them.
The price for dropping the assumption of \isaterm{f} being non-erasing here is a somewhat more complicated proof that may be sketched as follows:

\begin{proof}[sketch of proof of \isaterm{\isacommand{lemma} pushy{\isacharunderscore}{\kern0pt}strong{\isacharunderscore}{\kern0pt}shift}]
The implication from right to left is in fact easy.
We show the converse, i.e., assume that \isaterm{axiom} is pushy with respect to \isaterm{f}.

Assume first that the maximum letter image length is greater than $1$.
Let $\Psi = \isaterm{f{\isacharcircum}{\kern0pt}{\isacharcircum}{\kern0pt}(p + 1)}$.
For contradiction assume that \isaterm{axiom} is not pushy with respect to $\Psi$.
It follows that there exists a constant $\ell$ such that any bounded factor of $\Psi^i(\isaterm{axiom})$ is shorter than $\ell$.
As \isaterm{axiom} is pushy with respect to \isaterm{f}, there exists a long enough bounded factor such that its preimage is some $\Psi^i(\isaterm{axiom})$ must be longer than $\ell$.
Here, long enough is based on the maximum letter image length of $\isaterm{f}$.
As a preimage of a bounded factor is bounded, we arrive at a contradiction.

The remaining case when the maximum letter image length is not greater than $1$ is easy.
\end{proof}

Note that the remaining assumption on the finiteness of the alphabet may not be dropped.
Indeed, consider the following example tailored to make the claim fail. 
Assume $p > 0$, and consider the morphism $\varphi$ determined by
\[
b_i \mapsto \begin{cases}
b_{i+1}^{i+1} & \text{if $i+1 \neq k(p+1)$;} \\
\varepsilon & \text{ otherwise},
\end{cases}
\quad \quad
c_i \mapsto \begin{cases}
c_{i+1} & \text{if $i \neq k(p+1)$;} \\
b_{i+1}^{(i+1)!} c_{i+1} & \text{ otherwise},
\end{cases}
\]
The claim does not hold if the axiom equals $c_0$ since we obtain
\[
\varphi^i(c_0) = \begin{cases}
c_i & \text{ if $i = k(p+1)$;} \\
b_i^{i!} c_i & \text{ otherwise}.
\end{cases}
\]
As $\varphi^{p+1}(b_i) = \varepsilon$, each letter $b_i$ is bounded, hence the axiom $c_0$ is pushy with respect to $\varphi$.
On the other hand, the letters $c_i$ are not bounded (with respect to $\varphi$, which is equivalent\footnote{This claim is formalized under the name \isaterm{bounded\_pow\_bounded}.} to be with respect to $\varphi^{p+1}$).
As $\left(\varphi^{p+1}\right)^i(c_0) = c_{i(p+1)}$, the axiom $c_0$ is not pushy with respect to $\varphi^{p+1}$.

A \emph{morphic} word $\vv$ is a morphic image, by a morphism $\Psi$, of a purely morphic word, i.e., $\Psi(\varphi^k(a))$ constitute longer and longer prefixes of the word $\vv$, provided that $\lim_{k \to +\infty}|\Psi(\varphi^k(a))| = +\infty$.
We thus define a morphic language as follows:

\begin{isaframe}
\isacommand{definition}\isamarkupfalse%
\ morphic{\isacharunderscore}{\kern0pt}language\ {\isacharcolon}{\kern0pt}{\isacharcolon}{\kern0pt}\ {\isachardoublequoteopen}{\isacharparenleft}{\kern0pt}{\isacharprime}{\kern0pt}a\ list\ {\isasymRightarrow}\ {\isacharprime}{\kern0pt}a\ list{\isacharparenright}{\kern0pt}\ {\isasymRightarrow}\ {\isacharparenleft}{\kern0pt}{\isacharprime}{\kern0pt}a\ list\ {\isasymRightarrow}\ {\isacharprime}{\kern0pt}b\ list{\isacharparenright}{\kern0pt}\ {\isasymRightarrow}\ {\isacharprime}{\kern0pt}a\ list\ {\isasymRightarrow}\ {\isacharprime}{\kern0pt}b\ list\ set{\isachardoublequoteclose}\isanewline
\ \ \isakeyword{where}\ {\isachardoublequoteopen}morphic{\isacharunderscore}{\kern0pt}language\ f\ h\ axiom\ {\isasymequiv}\ factor{\isacharunderscore}{\kern0pt}closure\ {\isacharparenleft}{\kern0pt}h{\isacharbackquote}{\kern0pt}purely{\isacharunderscore}{\kern0pt}morphic{\isacharunderscore}{\kern0pt}language\ f\ axiom{\isacharparenright}{\kern0pt}{\isachardoublequoteclose}
\end{isaframe}

The variable \isaterm{h} represents the outer morphism, and the language is simply the factor closure of the image by \isaterm{h} of \isaterm{purely{\isacharunderscore}{\kern0pt}morphic{\isacharunderscore}{\kern0pt}language\ f\ axiom}.
An inductive definition would also be possible, but that would not allow for a direct use of facts on the underlying purely morphic language.
As a reuse of existing facts in formalization is rather preferred, the direct, less powerful, definition was chosen and specific useful claims are proven explicitly in the formalization.
One such claim is the following induction rule with fixed endomorphism \isaterm{f} and morphism \isaterm{h}:
\begin{isaframe}
\isacommand{lemma}\isamarkupfalse%
\ morphic{\isacharunderscore}{\kern0pt}language{\isacharunderscore}{\kern0pt}induct{\isacharbrackleft}{\kern0pt}consumes\ {\isadigit{1}}{\isacharbrackright}{\kern0pt}{\isacharcolon}{\kern0pt}\ \isakeyword{assumes}\isanewline
\ \ \ \ {\isachardoublequoteopen}x\ {\isasymin}\ morphic{\isacharunderscore}{\kern0pt}language\ f\ h\ axiom{\isachardoublequoteclose}\isanewline
\ \ \ \ {\isachardoublequoteopen}P\ {\isacharparenleft}{\kern0pt}h\ axiom{\isacharparenright}{\kern0pt}{\isachardoublequoteclose}\ \isakeyword{and}\isanewline
\ \ \ \ ih{\isadigit{2}}{\isacharcolon}{\kern0pt}\ {\isachardoublequoteopen}{\isasymAnd}w{\isachardot}{\kern0pt}\ h\ w\ {\isasymin}\ morphic{\isacharunderscore}{\kern0pt}language\ f\ h\ axiom\ {\isasymLongrightarrow}\ P\ {\isacharparenleft}{\kern0pt}h\ w{\isacharparenright}{\kern0pt}\ {\isasymLongrightarrow}\ P\ {\isacharparenleft}{\kern0pt}h\ {\isacharparenleft}{\kern0pt}f\ w{\isacharparenright}{\kern0pt}{\isacharparenright}{\kern0pt}{\isachardoublequoteclose}\ \isakeyword{and}\isanewline
\ \ \ \ ih{\isadigit{3}}{\isacharcolon}{\kern0pt}\ {\isachardoublequoteopen}{\isasymAnd}w\ u{\isachardot}{\kern0pt}\ w\ {\isasymin}\ morphic{\isacharunderscore}{\kern0pt}language\ f\ h\ axiom\ {\isasymLongrightarrow}\ P\ w\ {\isasymLongrightarrow}\ u\ {\isasymle}f\ w\ {\isasymLongrightarrow}\ P\ u{\isachardoublequoteclose}\isanewline
\ \ \ \ \isakeyword{shows}\ {\isachardoublequoteopen}P\ x{\isachardoublequoteclose}
\end{isaframe}

\section{Infinite words} \label{sec:infinite_words}

Similarly to the natural choice of the datatype for a finite word, an infinite word can be directly represented by the type 
\astream{}, defined in the theory ``HOL-Library.Stream'' from the main distribution of Isabelle.
In this theory, the type \astream{} is defined as a (corecursive) codatatype as follows:


\begin{isaframe}
\isacommand{codatatype}\isamarkupfalse%
\ {\isacharparenleft}{\kern0pt}sset{\isacharcolon}{\kern0pt}\ {\isacharprime}{\kern0pt}a{\isacharparenright}{\kern0pt}\ stream\ {\isacharequal}{\kern0pt}\isanewline
\ \ SCons\ {\isacharparenleft}{\kern0pt}shd{\isacharcolon}{\kern0pt}\ {\isacharprime}{\kern0pt}a{\isacharparenright}{\kern0pt}\ {\isacharparenleft}{\kern0pt}stl{\isacharcolon}{\kern0pt}\ {\isachardoublequoteopen}{\isacharprime}{\kern0pt}a\ stream{\isachardoublequoteclose}{\isacharparenright}{\kern0pt}\ {\isacharparenleft}{\kern0pt}\isakeyword{infixr}\ {\isacartoucheopen}{\isacharhash}{\kern0pt}{\isacharhash}{\kern0pt}{\isacartoucheclose}\ {\isadigit{6}}{\isadigit{5}}{\isacharparenright}{\kern0pt}
\end{isaframe}

The type is defined via 2 destructors. 
The first one is \isaterm{shd}, standing for ``stream head'', of type \isaterm{{\isacharprime}{\kern0pt}a}, representing the first element of the infinite sequence;
the second is \isaterm{stl}, standing for ``stream tail'', of type \astream{}, representing the rest of the sequence when the head is removed.
This useful definition does much more than to just declare the 2 destructors.
For instance, it also defines the constructor \isaterm{SCons}, with the notation \isaterm{{\isacharhash}{\kern0pt}{\isacharhash}}, relating the the two destructors: \isaterm{a {\isacharhash}{\kern0pt}{\isacharhash} uu} is the infinite sequence with head equal to \isaterm{a} and tail to \isaterm{uu}.



In the context of combinatorics on words, the following primitively recursive function (of ``HOL-Library.Stream'') is of interest as it prepends a finite word to an infinite word:

\begin{isaframe}
\isacommand{primrec}\isamarkupfalse%
\ shift\ {\isacharcolon}{\kern0pt}{\isacharcolon}{\kern0pt}\ {\isachardoublequoteopen}{\isacharprime}{\kern0pt}a\ list\ {\isasymRightarrow}\ {\isacharprime}{\kern0pt}a\ stream\ {\isasymRightarrow}\ {\isacharprime}{\kern0pt}a\ stream{\isachardoublequoteclose}\ \ \isakeyword{where}\isanewline
\ \ {\isachardoublequoteopen}shift\ {\isacharbrackleft}{\kern0pt}{\isacharbrackright}{\kern0pt}\ s\ {\isacharequal}{\kern0pt}\ s{\isachardoublequoteclose}\isanewline
{\isacharbar}{\kern0pt}\ {\isachardoublequoteopen}shift\ {\isacharparenleft}{\kern0pt}x\ {\isacharhash}{\kern0pt}\ xs{\isacharparenright}{\kern0pt}\ s\ {\isacharequal}{\kern0pt}\ x\ {\isacharhash}{\kern0pt}{\isacharhash}{\kern0pt}\ shift\ xs\ s{\isachardoublequoteclose}
\end{isaframe}

It is defined on the 2 constructors of \alist{}: the empty word \isaterm{{\isacharbrackleft}{\kern0pt}{\isacharbrackright}}, and \isaterm{x\ {\isacharhash}{\kern0pt}\ xs}, which stands for the prepending of the letter \isaterm{a} to the word \isaterm{xs} (which equals to the concatenation of the singleton \isaterm{[a]} to the word \isaterm{xs}).
In the formalization, we adopt the notation \isaterm{\isactrlbold {\isasymcdot}} for the mapping \isaterm{shift}.
Note that it is a bold cdot symbol, compared to the non-bold one for the concatenation of two finite words.
This choice is done even though Isabelle allows for overloading of a notation.
The reason is that the overloading prevents automatic type inference in some cases, which results in the need to explicitly specify the datatypes of variables often, which in turn makes the claims somewhat less readable.

Prefix and factor of an infinite word are defined in a straightforward manner:

\begin{isaframe}
\isacommand{definition}\isamarkupfalse%
\ sprefix{\isacharcolon}{\kern0pt}{\isacharcolon}{\kern0pt}\ {\isachardoublequoteopen}{\isacharprime}{\kern0pt}a\ list\ {\isasymRightarrow}\ {\isacharprime}{\kern0pt}a\ stream\ {\isasymRightarrow}\ bool{\isachardoublequoteclose}\isanewline
\ \ \isakeyword{where}\ {\isachardoublequoteopen}sprefix\ p\ uu\ {\isasymequiv}\ {\isacharparenleft}{\kern0pt}{\isasymexists}vv{\isachardot}{\kern0pt}\ uu\ {\isacharequal}{\kern0pt}\ p\isactrlbold {\isasymcdot}vv{\isacharparenright}{\kern0pt}{\isachardoublequoteclose}
\end{isaframe}

\begin{isaframe}
\isacommand{definition}\isamarkupfalse%
\ sfactor{\isacharcolon}{\kern0pt}{\isacharcolon}{\kern0pt}\ {\isachardoublequoteopen}{\isacharprime}{\kern0pt}a\ list\ {\isasymRightarrow}\ {\isacharprime}{\kern0pt}a\ stream\ {\isasymRightarrow}\ bool{\isachardoublequoteclose}\isanewline
\ \ \isakeyword{where}\ {\isachardoublequoteopen}sfactor\ w\ uu\ {\isasymequiv}\ {\isacharparenleft}{\kern0pt}{\isasymexists}p\ vv{\isachardot}{\kern0pt}\ uu\ {\isacharequal}{\kern0pt}\ p\isactrlbold {\isasymcdot}w\isactrlbold {\isasymcdot}vv{\isacharparenright}{\kern0pt}{\isachardoublequoteclose}
\end{isaframe}

As the names ``prefix'' and ``factor'' are already used for prefix and factor of a finite word, the prefix s is used in accordance with the library ``HOL-Library.Stream'' (\isaterm{hd} is the first letter of a finite word and \isaterm{tl} is its tail).


The language of an infinite word is just a collection of all its factors.

\begin{isaframe}
\isacommand{definition}\isamarkupfalse%
\ word{\isacharunderscore}{\kern0pt}language\ {\isacharparenleft}{\kern0pt}{\isachardoublequoteopen}{\isasymF}{\isachardoublequoteclose}{\isacharparenright}{\kern0pt}\isanewline
\ \ \isakeyword{where}\ {\isachardoublequoteopen}word{\isacharunderscore}{\kern0pt}language\ uu\ {\isasymequiv}\ {\isacharbraceleft}{\kern0pt}w{\isachardot}{\kern0pt}\ sfactor\ w\ uu{\isacharbraceright}{\kern0pt}{\isachardoublequoteclose}
\end{isaframe}

We adopt the notation $\isaterm{\isasymF \ uu}$ for the language of an infinite word \isaterm{uu}.
Similarly, the definition of the \emph{factor complexity} of an infinite word, that is, the mapping counting the number of factors of given length, is also straightforward:

\begin{isaframe}
\isacommand{definition}\isamarkupfalse%
\ factor{\isacharunderscore}{\kern0pt}complexity\isanewline
\ \ \isakeyword{where}\ {\isachardoublequoteopen}factor{\isacharunderscore}{\kern0pt}complexity\ uu\ {\isasymequiv}\ {\isacharparenleft}{\kern0pt}{\isasymlambda}i{\isachardot}{\kern0pt}\ card\ {\isacharparenleft}{\kern0pt}{\isacharparenleft}{\kern0pt}{\isasymF}\ uu{\isacharparenright}{\kern0pt}\isactrlbsub i\isactrlesub {\isacharparenright}{\kern0pt}{\isacharparenright}{\kern0pt}{\isachardoublequoteclose}
\end{isaframe}

Here, \isaterm{L\isactrlbsub i\isactrlesub} stand for the elements of the language \isaterm{L} of length \isaterm{i}.
The term \isaterm{card} is the cardinality of a set.
The symbol \isaterm{\isasymlambda} introduces a lambda expression, which makes \isaterm{factor{\isacharunderscore}{\kern0pt}complexity} a mapping from natural numbers to natural numbers.


As of now, the adaptation of \astream{} for combinatorics on words was immediate.
The next concepts are slightly less obvious.
First, we want to produce an infinite word based on an expression for its $n$-th letter.
This is done by the following primitively corecursive function:

\begin{isaframe}
\isacommand{primcorec}\isamarkupfalse%
\ snth{\isacharunderscore}{\kern0pt}stream\ {\isacharcolon}{\kern0pt}{\isacharcolon}{\kern0pt}\ {\isachardoublequoteopen}{\isacharparenleft}{\kern0pt}nat\ {\isasymRightarrow}\ {\isacharprime}{\kern0pt}a{\isacharparenright}{\kern0pt}\ {\isasymRightarrow}\ {\isacharprime}{\kern0pt}a\ stream{\isachardoublequoteclose}\ \isakeyword{where}\isanewline
\ \ {\isachardoublequoteopen}shd\ {\isacharparenleft}{\kern0pt}snth{\isacharunderscore}{\kern0pt}stream\ s{\isacharparenright}{\kern0pt}\ {\isacharequal}{\kern0pt}\ {\isacharparenleft}{\kern0pt}s\ {\isadigit{0}}{\isacharparenright}{\kern0pt}{\isachardoublequoteclose}\isanewline
{\isacharbar}{\kern0pt}\ {\isachardoublequoteopen}stl\ {\isacharparenleft}{\kern0pt}snth{\isacharunderscore}{\kern0pt}stream\ s{\isacharparenright}{\kern0pt}\ {\isacharequal}{\kern0pt}\ snth{\isacharunderscore}{\kern0pt}stream\ {\isacharparenleft}{\kern0pt}{\isasymlambda}i{\isachardot}{\kern0pt}\ s\ {\isacharparenleft}{\kern0pt}Suc\ i{\isacharparenright}{\kern0pt}{\isacharparenright}{\kern0pt}{\isachardoublequoteclose}
\end{isaframe}

The argument \isaterm{s} of \isaterm{snth{\isacharunderscore}{\kern0pt}stream s} is of type \isaterm{nat\ {\isasymRightarrow}\ {\isacharprime}{\kern0pt}a}, that is, a mapping from natural number to the alphabet, representing the sequence of letters of the infinite word.
The corecursive function is defined on the 2 destructors of the infinite word, saying first that its head is $\isaterm{s 0}$ and its tail is defined using \isaterm{s} shifted by $1$ as \isaterm{Suc\ i} is the successor of \isaterm{i}, i.e., \isaterm{i+1}.

Another preferred method of producing new infinite words is applying a morphism to some infinite word.
This can be achieved by the following combination of existing tools of ``HOL-Library.Stream'' and the elementary combinatorics on words package:

\begin{isaframe}
\isacommand{definition}\isamarkupfalse%
\ morph{\isacharunderscore}{\kern0pt}stream{\isacharcolon}{\kern0pt}{\isacharcolon}{\kern0pt}{\isachardoublequoteopen}{\isacharparenleft}{\kern0pt}{\isacharprime}{\kern0pt}a\ list\ {\isasymRightarrow}\ {\isacharprime}{\kern0pt}b\ list{\isacharparenright}{\kern0pt}\ {\isasymRightarrow}\ {\isacharprime}{\kern0pt}a\ stream\ {\isasymRightarrow}\ {\isacharprime}{\kern0pt}b\ stream{\isachardoublequoteclose}\isanewline
\ \ \isakeyword{where}\ {\isachardoublequoteopen}morph{\isacharunderscore}{\kern0pt}stream\ f\ uu\ {\isasymequiv}\ flat\ {\isacharparenleft}{\kern0pt}sfilter\ {\isacharparenleft}{\kern0pt}{\isasymlambda}w{\isachardot}{\kern0pt}\ w\ {\isasymnoteq}\ {\isasymepsilon}{\isacharparenright}{\kern0pt}\ {\isacharparenleft}{\kern0pt}smap\ f\isactrlsup {\isasymC}\ uu{\isacharparenright}{\kern0pt}{\isacharparenright}{\kern0pt}{\isachardoublequoteclose}
\end{isaframe}

The mapping \isaterm{f} is of type \isaterm{a\ list\ {\isasymRightarrow}\ {\isacharprime}{\kern0pt}b\ list} and the definition in fact does not require for it to be a morphism for the same reasons as in the case of a purely morphic language above.
The term \isaterm{f\isactrlsup {\isasymC}} stands for the ``core'' of \isaterm{f}, which is the underlying mapping of letters (not words) to its images.
The mapping \isaterm{smap} applies \isaterm{f\isactrlsup {\isasymC}} to each element of \isaterm{uu} producing a sequence of words, more precisely a term of type \blistlist{}.
To concatenate this sequence of words into the desired word of type \blist{} it suffices to use the mapping \isaterm{flat}.
However, the flattened sequence is first pruned: the empty words in the sequence are ignored using the \isaterm{sfilter} function.
The reason for this step is that the definition of \isaterm{flat} does not work for empty words in the flattened sequence.

Therefore, the given definition thus results in what we usually mean by a mapping of an infinite word $\uu = u_0u_1u_2\dots$:
\[
\varphi(\uu) = \varphi(u_0)\varphi(u_1)\varphi(u_2)\dots
\]
with the marginal exception when $\varphi(\uu)$ would in fact be a finite word.
In that case, we would end up with a term \isaterm{morph{\isacharunderscore}{\kern0pt}stream\ f\ uu} for which no reasonable claim could be proven.

As one of the motivations to introduce \isaterm{morph{\isacharunderscore}{\kern0pt}stream} is to define the fixed point of a morphism, which usually requires some assumption to prevent this case of shrinking an infinite word to a finite one, the definition above is satisfactory.
More specifically, this assumption is that the morphism $\varphi$ is \emph{prolongeable} on the first letter of its fixed point, that is, we have
\begin{equation} \label{eq:growing}
\lim_{i \to +\infty} \left| \varphi^i (u_0) \right| = +\infty
\end{equation}
for $\uu = u_0u_1u_2\dots$ with $\varphi(\uu) = \uu$.
Such word $\uu$ is a \emph{fixed point} of the morphism $\varphi$.
This concept is formalized as a locale.

\begin{isaframe}
\isacommand{locale}\isamarkupfalse%
\ fixed{\isacharunderscore}{\kern0pt}point\ {\isacharequal}{\kern0pt}\ endomorphism\ {\isacharplus}{\kern0pt}\isanewline
\ \ \isakeyword{fixes}\ {\isasymuu}\isanewline
\ \ \isakeyword{assumes}\isanewline
\ \ fixed{\isacharunderscore}{\kern0pt}point{\isacharcolon}{\kern0pt}\ {\isachardoublequoteopen}{\isacharparenleft}{\kern0pt}morph{\isacharunderscore}{\kern0pt}stream\ f\ {\isasymuu}{\isacharparenright}{\kern0pt}\ {\isacharequal}{\kern0pt}\ {\isasymuu}{\isachardoublequoteclose}\ \isakeyword{and}\isanewline
\ \ shd{\isacharunderscore}{\kern0pt}grow{\isacharcolon}{\kern0pt}\ {\isachardoublequoteopen}growing\ f\ {\isacharbrackleft}{\kern0pt}shd\ {\isasymuu}{\isacharbrackright}{\kern0pt}{\isachardoublequoteclose}\ 
\end{isaframe}

The second assumption stands exactly for the requirement \eqref{eq:growing}.
Within this locale, we can indeed show that the language of the fixed point \isaterm{{\isasymF}\ {\isasymuu}} equals the purely morphic language of the morphism \isaterm{f} with the axiom equal to the word of length $1$ consisting of the first letter of \isaterm{\isasymuu}:

\begin{isaframe}
\isacommand{lemma}\isamarkupfalse%
\ wlan{\isacharunderscore}{\kern0pt}pmor{\isacharcolon}{\kern0pt}\ {\isachardoublequoteopen}{\isasymF}\ {\isasymuu}\ {\isacharequal}{\kern0pt}\ purely{\isacharunderscore}{\kern0pt}morphic{\isacharunderscore}{\kern0pt}language\ f\ {\isacharparenleft}{\kern0pt}{\isacharbrackleft}{\kern0pt}shd\ {\isasymuu}{\isacharbrackright}{\kern0pt}{\isacharparenright}{\kern0pt}{\isachardoublequoteclose}
\end{isaframe}

A purely periodic infinite word, i.e., a word of the form $uuu\dots$, is in fact already present in ``HOL-Library.Stream'' as \isaterm{cycle\ u}, where \isaterm{u} is the repetend.
We define this notion is a straightforward manner:

\begin{isaframe}
\isacommand{definition}\isamarkupfalse%
\ purely{\isacharunderscore}{\kern0pt}periodic\ \isakeyword{where}\isanewline
\ \ {\isachardoublequoteopen}purely{\isacharunderscore}{\kern0pt}periodic\ uu\ {\isasymequiv}\ {\isasymexists}u{\isachardot}{\kern0pt}\ uu\ {\isacharequal}{\kern0pt}\ cycle\ u{\isachardoublequoteclose}
\end{isaframe}

It is not difficult to prove that the factor complexity of a purely periodic word is eventually constant, and it is bounded by the number of conjugates of the repetend.
For this, we use the \emph{primitive root} mapping from Combinatorics\_Words.
Given a non-empty word $w$, it returns the word $\rho(w)$ which satisfies $w = (\rho(w))^k$ for maximal $k$.

\begin{isaframe}
\isacommand{lemma}\isamarkupfalse%
\ pperiodic{\isacharunderscore}{\kern0pt}fac{\isacharunderscore}{\kern0pt}comp{\isacharcolon}{\kern0pt}\ \isakeyword{assumes}\ {\isachardoublequoteopen}w\ {\isasymnoteq}\ {\isasymepsilon}{\isachardoublequoteclose}\isanewline
\ \ \isakeyword{shows}\ {\isachardoublequoteopen}factor{\isacharunderscore}{\kern0pt}complexity\ {\isacharparenleft}{\kern0pt}cycle\ w{\isacharparenright}{\kern0pt}\ n\ {\isasymle}\ \isactrlbold {\isacharbar}{\kern0pt}{\isasymrho}\ w\isactrlbold {\isacharbar}{\kern0pt}{\isachardoublequoteclose}\isanewline
\ \ \ \ \ \ \ \ {\isachardoublequoteopen}n\ {\isasymge}\ \isactrlbold {\isacharbar}{\kern0pt}{\isasymrho}\ w\isactrlbold {\isacharbar}{\kern0pt}\ {\isasymLongrightarrow}\ factor{\isacharunderscore}{\kern0pt}complexity\ {\isacharparenleft}{\kern0pt}cycle\ w{\isacharparenright}{\kern0pt}\ n\ {\isacharequal}{\kern0pt}\ \isactrlbold {\isacharbar}{\kern0pt}{\isasymrho}\ w\isactrlbold {\isacharbar}{\kern0pt}{\isachardoublequoteclose}
\end{isaframe}

Analogous definition and similar statements are done for eventually periodic infinite words.


\section{Sturmian words} \label{sec:sturmian_words}

Sturmian words are infinite words that form a very popular subject of study in combinatorics on words.
They were studied in the article of 1940 by Morse and Hedlund \cite{HeMo}, and they have many equivalent characterizations (c.f \cite{BaPeSta2}).
In this formalization, we choose a constructive characterization which allows to parametrize the set of all Sturmian words.
More precisely, we choose the concept of lower mechanical words with irrational slope.
First, the $n$-th element needs to be defined for given real parameters $\alpha$ and $\beta$:


\begin{isaframe}
\isacommand{definition}\isamarkupfalse%
\ lower{\isacharunderscore}{\kern0pt}mechanical{\isacharunderscore}{\kern0pt}word{\isacharunderscore}{\kern0pt}n\ {\isacharcolon}{\kern0pt}{\isacharcolon}{\kern0pt}\ {\isachardoublequoteopen}real\ {\isasymRightarrow}\ real\ {\isasymRightarrow}\ int\ {\isasymRightarrow}\ int{\isachardoublequoteclose}\ \ \isakeyword{where}\isanewline
\ \ {\isachardoublequoteopen}lower{\isacharunderscore}{\kern0pt}mechanical{\isacharunderscore}{\kern0pt}word{\isacharunderscore}{\kern0pt}n\ {\isasymalpha}\ {\isasymbeta}\ n\ {\isasymequiv}\ {\isasymlfloor}{\isacharparenleft}{\kern0pt}n{\isacharplus}{\kern0pt}{\isadigit{1}}{\isacharparenright}{\kern0pt}{\isacharasterisk}{\kern0pt}{\isasymalpha}{\isacharplus}{\kern0pt}{\isasymbeta}{\isasymrfloor}\ {\isacharminus}{\kern0pt}\ {\isasymlfloor}n{\isacharasterisk}{\kern0pt}{\isasymalpha}{\isacharplus}{\kern0pt}{\isasymbeta}{\isasymrfloor}{\isachardoublequoteclose}\
\end{isaframe}

The infinite word is then seamlessly produced using \isaterm{snth{\isacharunderscore}{\kern0pt}stream} mentioned above.

\begin{isaframe}
\isacommand{definition}\isamarkupfalse%
\ lower{\isacharunderscore}{\kern0pt}mechanical{\isacharunderscore}{\kern0pt}word\ {\isacharcolon}{\kern0pt}{\isacharcolon}{\kern0pt}\ {\isachardoublequoteopen}real\ {\isasymRightarrow}\ real\ {\isasymRightarrow}\ int\ stream{\isachardoublequoteclose}\ \ \isakeyword{where}\isanewline
\ \ {\isachardoublequoteopen}lower{\isacharunderscore}{\kern0pt}mechanical{\isacharunderscore}{\kern0pt}word\ {\isasymalpha}\ {\isasymbeta}\ {\isasymequiv}\ snth{\isacharunderscore}{\kern0pt}stream\ {\isacharparenleft}{\kern0pt}lower{\isacharunderscore}{\kern0pt}mechanical{\isacharunderscore}{\kern0pt}word{\isacharunderscore}{\kern0pt}n\ {\isasymalpha}\ {\isasymbeta}{\isacharparenright}{\kern0pt}{\isachardoublequoteclose}
\end{isaframe}

Real numbers are not part of the main Isabelle package and one must import the theory ``HOL.Real''.
It provides enough tools to manipulate real numbers to formalize (and prove) the following claims:

\begin{isaframe}
\isacommand{lemma}\isamarkupfalse%
\ lmw{\isacharunderscore}{\kern0pt}rat{\isacharunderscore}{\kern0pt}per{\isacharcolon}{\kern0pt}\ \isakeyword{assumes}\ {\isachardoublequoteopen}{\isasymalpha}\ {\isasymin}\ {\isasymrat}{\isachardoublequoteclose}\isanewline
\ \ \isakeyword{shows}\ {\isachardoublequoteopen}purely{\isacharunderscore}{\kern0pt}periodic\ {\isacharparenleft}{\kern0pt}lower{\isacharunderscore}{\kern0pt}mechanical{\isacharunderscore}{\kern0pt}word\ {\isasymalpha}\ {\isasymbeta}{\isacharparenright}{\kern0pt}{\isachardoublequoteclose}\
\isanewline \isanewline
\isacommand{lemma}\isamarkupfalse%
\ lmw{\isacharunderscore}{\kern0pt}per{\isacharunderscore}{\kern0pt}rat{\isacharcolon}{\kern0pt}\ \isakeyword{assumes}\ {\isachardoublequoteopen}periodic\ {\isacharparenleft}{\kern0pt}lower{\isacharunderscore}{\kern0pt}mechanical{\isacharunderscore}{\kern0pt}word\ {\isasymalpha}\ {\isasymbeta}{\isacharparenright}{\kern0pt}{\isachardoublequoteclose}\isanewline
\ \ \isakeyword{shows}\ {\isachardoublequoteopen}{\isasymalpha}\ {\isasymin}\ {\isasymrat}{\isachardoublequoteclose}
\end{isaframe}

The claims exhibit the well-known fact that if $\alpha$ is rational, the lower mechanical word with slope $\alpha$ is periodic and vice versa.
Hence, if $\alpha$ is irrational, the word \isaterm{lower{\isacharunderscore}{\kern0pt}mechanical{\isacharunderscore}{\kern0pt}word\ {\isasymalpha}\ {\isasymbeta}} is an aperiodic word which is a Sturmian word.

The proofs of the last two lemmas are done via rotation of the unit circle, which is used in fact in another known characterization of Sturmian words.

\section{Structure of the formalization}

The formalization presented in the previous 3 sections is available in the Combinatorics on Words Formalized project repository \cite{CoW_gitlab_v18} as part of the package ``CoW\_Infinite'' split into 3 theories:
\begin{enumerate}
\item \texttt{CoW\_Infinite/Languages.thy} covers languages in general and (purely) morphic languages, described in \Cref{sec:languages};
\item \texttt{CoW\_Infinite/Infinite\_Words.thy} covers infinite words, described in \Cref{sec:infinite_words};
\item \texttt{CoW\_Infinite/Sturmian\_Words.thy} covers lower mechanical words, described in \Cref{sec:sturmian_words}.
\end{enumerate}

While the first two theories are intended to serve as a general backbone of the extension of Combinatorics on Words Formalized project to infinite words, the last theory of Sturmian words is rather a proof of concept theory, to demonstrate that the backbone sessions work well on a popular matter.

\section*{Acknowledgements}

The author acknowledges support by the Czech Science Foundation grant GA\v CR 20-20621S.
The author also expresses his gratitude to the other authors of the Combinatorics on Words Formalized project, Štěpán Holub and Martin Raška.

\bibliographystyle{splncs04}
\bibliography{Words23}

\end{document}